\newcommand{\Pbb}{\mathbb{P}}
\newcommand{\Ebb}{\mathbb{E}}
\newcommand{\rmd}{\mathrm{d}}
\newcommand{\RE}{\operatorname{Re}}
\newcommand{\norm}[1]{\left\Vert#1\right\Vert}
\newcommand{\openone}{\mathds1}

\documentclass[twocolumn,showpacs,preprintnumbers,amsmath,amssymb]{revtex4}

\usepackage{graphicx}
\usepackage{dcolumn}
\usepackage{bm}
\usepackage{float}

\begin{document}

\title{Stochastic wave-function unravelling of the generalized Lindblad equation}

\author{V. Semin}

\affiliation{Samara National Research University, 34, Moskovskoe shosse, Samara, 443086, Russian Federation}
\email{semin@ssau.ru}

\author{I. Semina}

\affiliation{Quantum Research Group, School of Physics, University of KwaZulu-Natal,
Durban, 4001, South Africa} \email{yusov@list.ru}

\author{F. Petruccione}

\affiliation{Quantum Research Group, School of Chemistry and Physics,
University of KwaZulu-Natal, Durban 4001, South Africa
and National Institute for Theoretical Physics (NITheP),
KwaZulu-Natal, Durban 4001, South Africa} \email{petruccione@ukzn.ac.za}

\date{\today}

\begin{abstract}
We investigate generalized non-Markovian stochastic Schr\"{o}dinger equations (SSEs), driven by a multidimensional counting process and multidimensional Brownian motion introduced by A. Barchielli and C. Pellegrini [J. Math. Phys. 51, 112104 (2010)]. We show, that these SSEs can be translated in a non-linear form,  which can be efficiently simulated. The simulation is illustrated by the model of a two-level system in a structured bath and the results of the simulations are compared with the exact solution of the generalized master equation.

\end{abstract}


\maketitle
\section{Introduction}
The theory of open quantum systems is focused on the investigation of a system interacting with its environment\cite{4}. Usually, the time evolution of such systems can be found through the solution of the master equation for the reduced density matrix, typically in Lindblad form, that provides the complete positivity and trace preservation of the density operator \cite{12}. This approximation is based on the absence of memory effects in the system and has shown a good agreement with various experiments \cite{4,12,13, Nat1,Nat2,Nat3}. Alternatively, the dynamics of the open system can be described through the so-called stochastic Schr\"{o}dinger equations (SSEs) \cite{Bb}. The main idea of the method is to reproduce the solution of the master equation for the reduced density matrix through the ensemble average of a large number of realizations of the stochastic wave function \cite{NumMet}. To find the SSE corresponding to a given master equation is called \textit{unravelling}. Also in the Markovian case, the stochastic unravelling can be interpreted in terms of continuous measurements within the notions of the quantum trajectory theory.

In the non-Markovian case the systems may show strong and long memory effects. In this scenario, the master equation has non-Markovian form and to find the corresponding SSE, describing both the evolution of a quantum system and continuous monitoring, is a difficult task. However, a few strategies have been presented recently \cite{3,5,6, Request}. One such strategy is to start from the Markovian SSE and try to find the generalization for the non-Markovian case using colored noises and random coefficients \cite{Bb,6,Sem}. Another strategy is the unravelling of the generalized non-Markovian master equations \cite{1}. An alternative approach to introduce non-Markovianity based on the quantum stochastic calculus and quantum Langevin equation is presented in \cite{Request}.

There are several types of the SSEs appearing in the framework of the theory: diffusive, with jumps and combined, jump-diffusion SSEs. We will focus on the most general one, driven by multidimensional Brownian motion and multidimensional counting process.  The aim of the paper is to show, that the unravelling \cite{1} can be translated in a useful numerical approach. The simulations are demonstrated on the model of a two-level system in a structured bath. 

The article is organized as follows: in Section 2 we review the SSE corresponding to the generalized non-Markovian master equation of the Lindblad type; in Section 3 the general approach to the simulation of the particular type of the SSEs is presented; in Section 4 the SSE and the corresponding master equation are shown for the two-level system in a structured bath. Section 5 includes the simulation results for the SSE presented above. In section 6 we qualitatively compare our approach to some others. The last Section is the  conclusions.

\section{A generalized non-linear SSE}
Before starting the discussion let us remind some basis of the open quantum systems.
In the context of the theory of open quantum systems any quantum system is divided into two interacting subsystems, namely, the system of interest and its surrounding environment, called a thermostat. The thermostat usually has many degrees of freedom that are irrelevant and the main problem of the theory is to extract the information about the system of interest from the dynamics of the whole system. There are many approaches to do so and the most of them lead either to master equations or to SSEs.

In this section we will review two types of generalized non-Markovian equations, namely the master equation and the SSE \cite{1}, which play the main role in this article.

The first equation to be considered is the generalized non-Markovian master equation of the Lindblad-type \cite{Bud1,Br1,Br2,Bud2,Bud3}. In the complex Hilbert space $\mathcal{H}$ the initial condition of the reduced density operator $\eta_S(0)$ of the open system is characterized by the following properties:
\begin{equation}
\eta_S(t)=\sum_{i=1}^n\eta_i(t), \quad\eta_i(0)>0, \quad \sum_{i=1}^{n}\mathrm{tr_{\mathcal{H}}}\{\eta_i(0)\}=1,
\end{equation}
where the operator $\eta_i(t)$ corresponds to the projections of the reduced density operator onto subspaces in the Hilbert space of the thermostat.

For the vector $(\eta_1(t),\ldots,\eta_n(t))$, whose elements are trace class operators, the evolution equation has the form \cite{1}: 
\begin{eqnarray}\label{GME}
\frac{d}{dt}\eta_i(t)&=&-i[H^i,\eta_i(t)]\\
&+&\sum_{\alpha\in A}\sum^{n}_{k=1}\Big( R_\alpha^{ik}\eta_k(t)R_\alpha^{ik*}-\frac{1}{2}\left\{R_\alpha^{ik}R_\alpha^{ik*},\eta_k(t)\right\} \Big),\nonumber
\end{eqnarray}
 where $H^i,R^{ki}_\alpha$ are the system bounded operators, $\alpha\in A$ and $A$ is a finite set of indices.

For the particular choice of parameters in Eq. (\ref{GME}), such as: $A=\{-m_1,\ldots,-1,1,\ldots,m_2\}$, $R^{ij}_{-\alpha}=\delta_{ij}L^i_\alpha$, $\alpha=1,\ldots,m_1$, the master equation Eq. (\ref{GME}) takes the form: 
\begin{equation}\label{MME}
 \frac{d}{dt}\eta_i(t)= \mathcal{K}(\eta_1(t),\ldots,\eta_n(t)),
\end{equation} 
where
  \begin{eqnarray}
 &\mathcal{K}(\tau_1,\ldots,\tau_n):=-i[H^i,\tau_i]\\+
 \sum^{m_1}_{\alpha=1}&\left(L^i_\alpha\tau_i{L^{i}_\alpha}^*-\frac{1}{2}\{{L^{i}_\alpha}^* L_\alpha^i,\tau_i\}\right) \nonumber \\
+\sum_{\alpha=1}^{m_2}\sum_{k=1}^{n}&\left(R^{ik}_\alpha\tau_k {R^{ik}_\alpha}^*-\frac{1}{2}\{{R^{ki}_\alpha}^* R_\alpha^{ki},\tau_i\}\right).\hspace{-1.3cm} \nonumber
 \end{eqnarray}
 
  In the language of continuous measurement, the vector $\eta_i(t)$ is called the \textit{a priori state} and corresponds to the state of the system under the physical probability $\Pbb^T$ \cite{1}, when the output is not known: 
\begin{equation}\label{apriori}
\eta_i(t)=\Ebb_{\Pbb^T}\left[|\psi_i(t)\rangle\langle\psi_i(t)|\right],\quad i=1,\ldots,n, \quad T\geq t\geq 0,
\end{equation}
 where $\Ebb_{\Pbb^T}$ denotes the ensemble average over all possible realisations of the state vector $|\psi_i(t)\rangle$.
 
 The vector $\psi_i(t)$ is the conditional state of the system given the observed output up to time $t$, corresponding to the \textit{a posteriori state}. The equation for the vector $\psi_i(t)$ is the second equation to be considered and it is the generalized SSE \cite{1}:
\begin{eqnarray}\label{NMSSE}
\rmd\psi_j(t)=V_j\big(\psi_1(t_{-}),\ldots,\psi_n(t_{-}) \big)\rmd t\\
+\sum_{\alpha=1}^{d_1}\left(L^j_\alpha-\frac{1}{2}\upsilon_\alpha(t)\right)\psi_j(t_{-})\rmd\hat{W}_{\alpha}(t)\nonumber\\
+\sum^{d_2}_{\alpha=1}\sum^{n}_{k=1}\left(R^{jk}_{\alpha}\psi_k(t_{-})-\frac{1}{2}\upsilon^k_\alpha(t)\psi_j(t_{-})\right)\rmd\hat{W}^k_{\alpha}(t)\nonumber\\
+\sum_{\alpha=d_1+1}^{m_1}\left(\frac{L_\alpha^j}{\sqrt{I_{\alpha}(t)}}-1\right)\psi_j(t_{-})\rmd N_{\alpha}(t)\nonumber\\
+\sum^{m_2}_{\alpha=d_2+1}\sum^{n}_{k=1}\left(\frac{R^{jk}_{\alpha}\psi_k(t_{-})}{\sqrt{I^k_{\alpha}(t)}}-\psi_j(t_{-})\right)\rmd N^k_{\alpha}(t),\nonumber
\end{eqnarray}
where
\begin{eqnarray}\label{V}
V_j\big(\psi_1(t_{-}),\ldots,\psi_n(t_{-}) \big)=\nonumber\\
K^j\psi_j(t_{-})+\frac{1}{2}\sum^{m_1}_{\alpha=d_1+1}I_{\alpha}(t)\psi_j(t_{-})\nonumber\\
+\frac{1}{2}\sum^{m_2}_{\alpha=d_2+1}\sum^{n}_{k=1}I^k_{\alpha}(t)\psi_j(t_{-})\nonumber\\+
\frac{1}{2}\sum_{\alpha=1}^{d_1}\upsilon_{\alpha}(t)\left(L^j_\alpha-\frac{1}{4}\upsilon_\alpha(t)\right)\psi_j(t_{-})\nonumber\\
+\frac{1}{2}\sum^{d_2}_{\alpha=1}\sum^{n}_{k=1}\upsilon^k_{\alpha}\left(R^{jk}_\alpha\psi_k(t_{-})-\frac{1}{4}\upsilon^k_\alpha(t)\psi_j(t_{-})\right),
\end{eqnarray}
\begin{equation}
K^j=-iH^j-\frac{1}{2}\sum_{\alpha=1}^{m_1}{L^j_{\alpha}}^*L_\alpha^j-\frac{1}{2}\sum^{m_2}_{\alpha=1}\sum_{k=1}^{n}{R^{kj}_\alpha}^*R^{kj}_\alpha.
\end{equation}
Here $\psi_j(t_{-})$ shows the left limit of the process.
 
The processes $\hat{W}_\alpha(t),\hat{W}_\beta^k(t)$ are independent standard Wiener processes under the physical probability:
\begin{eqnarray}\label{Gir}
\hat{W}_\alpha(t)\!=\!W_\alpha(t)\!-\!\int\limits_{0}^{t}\!\!\upsilon_\alpha(s)ds,  \hat{W}_\beta^k(t)\!=\!W_\beta^k(t)\!-\!\int\limits_{0}^{t}\!\!\upsilon_\beta^k(s)ds,\\
t\in[0,T],\alpha=1,\ldots,d_1, \beta=1,\ldots,d_2,  k=1,\ldots,n
\end{eqnarray}
and the processes $N_\alpha(t),N_\beta^k(t)$ are the counting processes of the intensities $I_\alpha(t)$ and $I_\beta^k(t)$ with $\alpha=d_1+1,\ldots,m_1,\beta=d_2+1,\ldots,m_2,k=1,\ldots,n.$

The parameters $\upsilon_{\alpha}(t),\upsilon^k_{\alpha}(t),I_\alpha(t)$ and $I^k_\alpha(t)$ are
\begin{eqnarray}
\upsilon_{\alpha}(t)=2\RE\Big\langle\psi_j(t_{-})\Big | L^j_\alpha\psi_j(t_{-})\Big\rangle,\label{v}\\ \upsilon^k_{\alpha}(t)=2\RE\Big\langle\psi_j(t_{-})\Big | R^{jk}_\alpha\psi_k(t_{-})\Big\rangle, \label{v1}
\end{eqnarray}
\begin{equation}\label{I}
I_\alpha(t)=\sum^{n}_{j=1}\left\Vert L^j_\alpha\psi_j(t_{-})\right\Vert^2, I^k_\alpha(t)=\sum^{n}_{j=1}\left\Vert R^{jk}_\alpha\psi_j(t_{-})\right\Vert^2.
\end{equation}

Eq. \eqref{NMSSE} is a non-linear SSE, due to the nonlinearity of the terms Eq. (\ref{v}), Eq. (\ref{v1}) and Eq. (\ref{I}). The equation can be the starting point for the numerical simulations. Moreover, the driving noises $N^k_\alpha$ are state-dependent through the stochastic intensities $I_\alpha$ and $I^k_\alpha$. In the next section we will propose the general ideas on the simulation of Eq. \eqref{NMSSE}.

\section{The jump-adapted scheme}
Markovian and non-Markovian SSEs can be an instrument for the numerical simulations through the stochastic wave function method \cite{NumMet, 2,Sem}.  Here we introduce the general ideas for the simulation of a jump-diffusion SSE Eq. (\ref{NMSSE}) in order to describe the evolution of the concrete physical system, presented in the next section. It is worth noting that the main difference from the case presented in \cite{Merv} is taking into account the diffusive terms. Moreover, the appearance of such terms induces the time-dependent rates in the simulations. This fact leads to the generalization presented below. For a generic numerical treatment of a jump-diffusion stochastic differential equations see \cite{Pl}.

Let us rewrite the jump-diffusion SSE in the following form:
\begin{eqnarray}\label{JDS}
d\psi(t)&=&
a(\psi(t))dt\\&+&\sum_{i=1}^{k}b_i(\psi(t))dW_i(t)+\sum_{j=1}^{m}c_j(\psi(t))dN_j(t),\nonumber\\ &0\leq & \!\!t\leq T, \nonumber
\end{eqnarray}
where $a(\psi(t)), b_i(\psi(t))$ and $c_j(\psi(t))$ are the vector functions, $W_i$ are the standard Wiener processes and $N_j$ are the counting processes with the stochastic intensities: 
\begin{equation}\label{inens}
I_j=f_j(\psi(t)).
\end{equation}
The presence of these state-dependent intensities leads to the deduction, that  the jumps  cannot be determined separately from the diffusion. This fact leads to the increasing of the cost of the simulations \cite{Glasserman}.  

In general, Eq. (\ref{JDS}) is not solvable due to the fact, that the equation for the mean value $\Ebb [F(\psi(T))]$ does not exist in closed form. Principally, this evaluation can be obtained through the numerical integration, which requires the distribution for $\psi(T)$. This leads to stochastic simulations as the general way to investigate Eq. (\ref{JDS}). The large number of realizations of the discrete approach of $\psi$ is generated numerically, while the average $\Ebb [F(\psi(T))],$ where $F$ is some functional of the wave-vector, is calculated from all the paths. The time discretization is performed on the time interval $[0,T]$:
\begin{equation}
0=t_1<t_2<\ldots<t_n=(n-1)\Delta t<\ldots<t_M=T,
\end{equation}
and $\Delta t=\frac{T}{M-1}$.

The simulation scheme for Eq. (\ref{JDS}) can be described by the following steps:

\begin{enumerate}

\item Select the corresponding initial values $\psi_0=\psi(t_0)$ from the initial distribution.

\item Take two random variables, uniformly distributed over the interval $[0,1]$. One variable will be used as the random waiting time $\tau$ and the second variable for the choice of the type of the count.

\item Calculate the initial values for the $I^k_\beta(\psi_0,t_0),\upsilon^k_\beta(\psi_0,t_0),V_j(\psi_0,t_0)$ with the help of Eqs. (\ref{V}),(\ref{v}),(\ref{I}).
Also, these quantities should be calculated again for any new $\psi(t)$: $I^k_\beta(\psi,t), \upsilon^k_\beta(\psi,t),V_j(\psi,t)$.

\item
In the absence of counts, the trajectory is purely diffusive and can be simulated using the standard discretization methods, such as the Euler scheme:
\begin{equation}\label{Euler}
\psi(t_{n+1})=\psi(t_n)+a\hspace{0.5mm}\psi(t_n)\Delta t+\sum^k_{i=1}b_i\hspace{0.5mm}\psi(t_n)\Delta W_i.
\end{equation}
\item The conditional probability of having the next count at time $t$, while the previous count was at time $t_0$ is exp$\{-\int_{t_0}^t  I(s)ds\}$, where $I(t)=\sum_{k,\beta}I^k_\beta(t)$ is the intensity of the number of counts of any type $N(t)=\sum_{k,\beta}N^k_\beta(t)$. Hence, if $\tau=$exp$\{-\int_{t_0}^t  I(s)ds\}$, there is a jump to a new state with the conditional probability $I^k_\beta(t)/I(t)$, which is also defined a type of the count. 

\item  Repeat steps 2-4 until the final time $T$ of interval $[0,T]$ is reached.

\item After a sufficiently large sample of all realizations is generated, the quantity of interest can be found through the corresponding averages.

\end{enumerate}

The resulting quantities can be found by averaging the standard formulas for the mean and deviation.  More precisely, the estimator for the mean value can be found by the expression for sample mean:
\begin{equation}\label{M}
\widehat{M}_T=\frac 1 R \sum^{R}_{r=1}F[\psi^r,T],
\end{equation}
where $R$ denotes the sample size and $r$ indicates the different realisations. The quantity of interest is a real functional $F[\psi,T]$ of the \textit{a posteriori states} $\psi(t),t\in[0,T]$.  

Let us discuss some insides of the above mentioned scheme. First of all, the central element of the scheme is the Euler approximation \eqref{Euler} that has the first order of weak convergence. At the same time the higher-order numerical schemes is much more complicated (see, for example, \cite{Pl}), especially in the case with more than one diffusive component. The implementation of such schemes is not an easy task and requires some additional work to evaluate the multidimensional stochastic integrals at each step \cite{Pl2}. Thus, the Euler scheme is the easiest for numerical realization method to deals with stochastic equations. Note that the Euler scheme may be replaced by some more sophisticated methods such as  Platen's or predictor-corrector approximations \cite{Pl2}.
Second, some problems with numerical stability may appear during the simulation, especially at sufficiently large time steps. This problem can be solved by decreasing of the time step. In any case monitoring of the norm during the computation allows to catch and exclude from consideration unstable trajectories of the process, when the norm is too large. The other steps of the suggested algorithm are pretty standard and are used quite often \cite{4}. The only significant difference is the explicit dependence of the intensities of count processes on a current stochastic trajectory of the simulated process. This dependence is efficiently controlled in the fifth step of the algorithm by correction of the intensity at each step.

In the next section we apply the suggested numerical scheme to a concrete physical model, namely a two-level system in a structured bath in order to study the influence of the diffusion components.  The results of the simulations are supported by the solution of generalized Lindblad rate equation Eq. (\ref{MME}).

\section{A two-level system in a structured bath}

In this section we consider a concrete physical model, corresponding to the generalized non-Markovian SSE Eq. (\ref{NMSSE}). This model has been proposed in \cite{1} and different special cases were studied in \cite{Br1,Br2,Merv,Merv1}. 

In \cite{Merv} the jumps with the state-independent intensities have been discussed. Unlike \cite{Merv}, here we study the influence of diffusion components and of jumps with state-dependent intensities. 

\begin{figure}
\begin{center}
\includegraphics[scale=0.46]{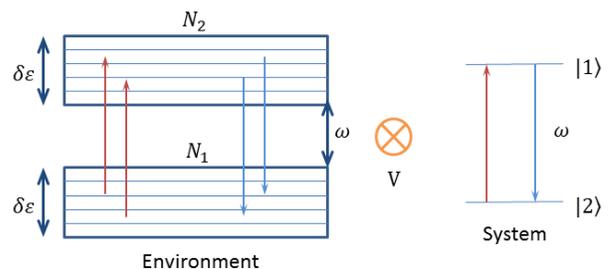}
\caption{A two-state system, with level distance $\omega$, couples to an environment consisting of two energy bands, each with finite number of evenly spaced levels $N_1$ and $N_2$. $\delta\epsilon$ is the width of the bands and $V$ is the system-environment interaction potential. }\label{mod0}
\end{center}
\end{figure}

 The aforementioned model is a two-level system with the ground state $|2\rangle$, the exited state $|1\rangle$, with the transition frequency $\omega$, coupled to a structured environment (see Fig. \ref{mod0}) \cite{Merv1,Br1}. 
 Such a model corresponds, for example, to a two-level atom injected in a photonic crystal with a photonic band gap.
 The environment consists of a large number of energy levels arranged in two energy bands of width $\delta\epsilon$, each with a finite number of equally spaced levels $N_1$ and $N_2$, divided by the gap $\omega$. The system-environment potential $V$ gives the overall strength of the interaction \cite{Merv1}
 \begin{equation}
V(n_1,n_2)=\lambda\sum_{n_1,n_2}c(n_1,n_2)\sigma_{+}|n_1\rangle\langle n_2|+\mathrm{H.c.}
 \end{equation}
where $n_1$ and $n_2$ are the levels of the lower and upper energy band. The parameter $\lambda$ represents the strength of the interaction and the constants $c(n_1,n_2)$ are complex Gaussian random variables with zero mean and unit variance. 

The total Hamiltonian, characterizing the quantum system \cite{Br3} is
\begin{eqnarray}
H=\frac{1}{2}\omega\sigma_z+\sum_{n_1}\frac{\delta\epsilon}{N_1}n_1 |n_1\rangle\langle n_1|\\+
\sum_{n_2}\left(\omega+\frac{\delta\epsilon}{N_2}n_2\right)|n_2\rangle\langle n_2|\hspace{5mm}+V(n_1,n_2)\nonumber.
\end{eqnarray}
The SSE for this model, derived in \cite{1} reads
\begin{widetext}
\begin{equation}\label{SSE}
\begin{cases}
 d\psi_1(t)=V_1\hspace{0.5mm}(\psi_1(t_{-}),\psi_2(t_{-}))\hspace{0.5mm}dt - \psi_1(t_{-})\left(dN_1^1(t)+dN_2^1(t)\right)  +\Big(\frac{\sigma_{+}\psi_2(t_{-})}{\norm{\sigma_{+}\psi_2(t_{-})}}-\psi_1(t_{-})\Big)\hspace{0.5mm}dN_1^2(t)\\ \hspace{0.5cm}+\sqrt{\gamma_0}\left(\sigma_{-}\hspace{0.5mm}\psi_1(t_{-})-\frac{1}{2}\upsilon(t)\hspace{0.5mm}\psi_1(t_{-})\right)\left(\sqrt{\epsilon}\hspace{0.5mm}d{W}_1(t)+\sqrt{1-\epsilon}\hspace{0.5mm}d{W}_2(t)\right), \vspace{0.5 cm}\\
d\psi_2(t)=V_2\hspace{0.5mm}(\psi_1(t_{-}),\psi_2(t_{-}))\hspace{0.5mm}dt - \left( \frac{\psi_1(t_{-})}{\norm{\psi_1(t_{-})}} - \psi_2(t_{-})\right)\hspace{0.5mm}dN_2^1(t)+\Big(\frac{\sigma_{-}\psi_1(t_{-})}{\norm{\sigma_{-}\psi_1(t_{-})}}-\psi_2(t_{-})\Big)\hspace{0.5mm}dN_1^1(t) - \psi_2(t_{-})\hspace{0.5mm}dN_1^2(t)
\\ \hspace{0.5cm}+\sqrt{\gamma_0}\left(\sigma_{-}\hspace{0.5mm}\psi_1(t_{-})-\frac{1}{2}\upsilon(t)\hspace{0.5mm}\psi_1(t_{-})\right)\left(\sqrt{\epsilon}\hspace{0.5mm}d{W}_1(t)+\sqrt{1-\epsilon}\hspace{0.5mm}dt{W}_2(t)\right),
\end{cases}
\end{equation}
\end{widetext}
where $W_i$ are the standard independent Wiener processes representing the emitted light, divided into two channels: channel 1 ($dW_1$) contains the light reaching the heterodyne detector and channel 2 ($dW_2$) contains the losses of the light. The processes $N_1^1,N_1^2,N^1_2$  are the counting independent  processes with probabilities $I_1^1,I_1^2,I_2^1$ and
\begin{eqnarray}
\upsilon(t)=2\sum_{k=1}^{2}\RE\langle\psi_k(t_{-})|\sigma_{-}\hspace{0.2mm}\psi_k(t_{-})\rangle,\\ I_2^1(t)=\varkappa\gamma_0\norm{\psi_1(t_{-})}^2,\\
I_1^1(t)=\gamma_1\norm{\sigma_{-}\hspace{0.2mm}\psi_1(t_{-})}^2,\\
I_1^2(t)=\gamma_2\norm{\sigma_{+}\hspace{0.2mm}\psi_2(t_{-})}^2,\\
\nonumber\\
V_j(\psi_1(t_{-}),\psi_2(t_{-}))=K^j\hspace{0.2mm}\psi_j(t_{-})\nonumber\\+\frac{I_1^1(t)+I_2^1(t)+I_1^2(t)}{2}\hspace{0.5mm}\psi_j(t_{-})\nonumber\\
+\frac{\gamma_0}{2}\upsilon(t)\sigma_{-}\hspace{0.5mm}\psi_j(t_{-}) - \frac{\gamma_0}{8}\upsilon(t)^2\psi_j(t_{-}),
\end{eqnarray}
where $\sigma_z, \sigma_{\pm}$ are the usual Pauli matrices.

The coefficients $K^j$ correspond to
\begin{eqnarray}
K^1=-\frac{i\omega_1}{2}\sigma_z - \frac{\gamma_0+\gamma_1}{2}P_{+}-\frac{\gamma_0 \varkappa}{2}\openone,\nonumber\\ K^2=-\frac{i\omega_2}{2}\sigma_z - \frac{\gamma_0}{2}P_{+}-\frac{\gamma_2}{2}P_{-}.\nonumber
\end{eqnarray}
$P_{+}=\sigma_{+}\sigma_{-}$ is the projection on the excited state $|1\rangle$ and $P_{-}=\sigma_{-}\sigma_{+}$ is the projection on the ground state $|2\rangle$. In these equations the terms with $\gamma_1$ and $\gamma_2$ correspond to the molecular transitions due to the environment accompanied with the transitions between the two bands of the environment. The term $\gamma_0$ is the explicit spontaneous emission and the term $\gamma_0\varkappa$ is a thermal-like excitation. The parameters $\omega_1$ and $\omega_2$ are the two energy shifts caused by the two bands of the environment. In this work we put $\omega_1=\omega_2$.
 
The model is defined by the following choices of parameters:
\begin{eqnarray}
\omega_i>0, \qquad \gamma_i>0,\qquad i=1,2;\nonumber\\
\varkappa>0,\qquad \gamma_0>0,\qquad 0<\epsilon\leq 1.
\end{eqnarray}

For this system the generalized Lindblad rate equation Eq. (\ref{MME}) takes the form \cite{1}:
\begin{equation}\label{Lindblad}
\begin{cases}
\frac{d}{dt}\eta_1(t)=\gamma_0\Bigl(\sigma_{-}\eta_1(t)\sigma_{+}-\frac{1}{2}\{P_{+},\eta_1(t)\}\Bigl)\\
\hspace{1cm}+\hspace{0.5mm}\gamma_2\sigma_{+}\eta_2(t)\sigma_{-}-\frac{\gamma_1}{2}\{P_{+},\eta_1(t)\}\\
-\gamma_0\varkappa\eta_1(t)-\frac{i\omega_1}{2}[\sigma_z,\eta_1(t)],\vspace{0.5 cm}\\
\frac{d}{dt}\eta_2(t)=\gamma_0\Bigl(\sigma_{-}\eta_2(t)\sigma_{+}-\frac{1}{2}\{P_{+},\eta_2(t)\}\Bigl)\\
\hspace{1cm}+\hspace{0.5mm}\gamma_1\sigma_{-}\eta_1(t)\sigma_{+}-\frac{\gamma_2}{2}\{P_{-},\eta_2(t)\}\\
+\gamma_0\varkappa\eta_1(t)-\frac{i\omega_2}{2}[\sigma_z,\eta_2(t)],
\end{cases}
\end{equation}
It should be mentioned, that if we have four states, $1^+,1^-,2+$ and $2^-$, the non-zero transition rates are $\gamma_0$ for the transition $1^+\rightarrow 1^-$, $\gamma_1$ for the transition $1^+\rightarrow 2^-$, $\gamma_0\varkappa$ - for $1^+\rightarrow 2^+$, $\gamma_2$ - for $2^-\rightarrow 1^+$, $\gamma_0$ - for $2^+\rightarrow 2^-$ and $\gamma_0\varkappa$ - for $1^-\rightarrow 2^-$.

\section{The stochastic simulation}
\begin{figure}[!h]
  \begin{center}
  \includegraphics[scale=0.5]{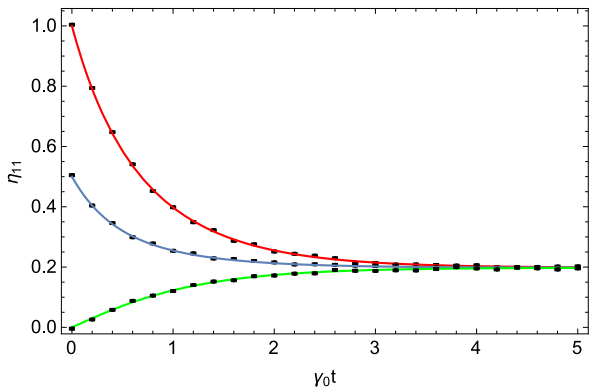}
  \caption{The ground state population for different initial values: red line - $\psi_1(0)=\psi_2(0)=\{\frac{1}{\sqrt{2}},0\}$; blue line - $\psi_1(0)=\{\frac{1}{\sqrt{2}},0\}$ and $\psi_2(0)=\{0,\frac{1}{\sqrt{2}}\}$; green line - $\psi_1(0)=\{0,\frac{1}{\sqrt{2}}\}$ and $\psi_2(0)=\{0,\frac{1}{\sqrt{2}}\}$. The parameters are: $\gamma_0=1, \gamma_1=0.5, \gamma_2=0.3, \omega_1=\omega_2=\sqrt{37}/2, \varkappa=2, \Delta t=0.001$. The error bars have the same size as the dots on the figure.}\label{InStates}
  \end{center}
  \end{figure}

\begin{figure}
 \begin{center}
 \includegraphics[scale=0.5]{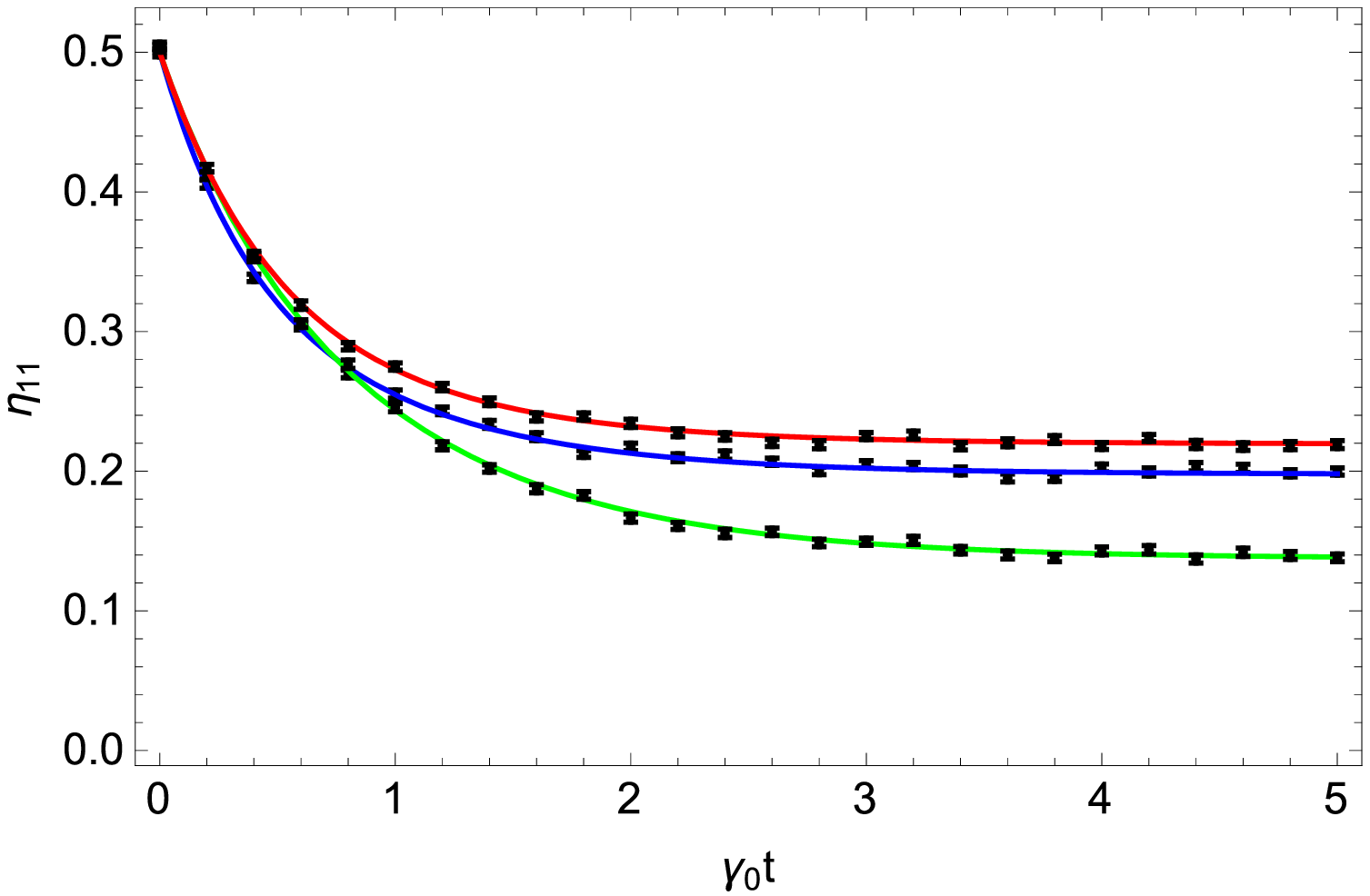}
 \caption{Plot of the mean occupation number of the excited state for different values of $\gamma_1$: red line - $\gamma_1=0.05$, blue line - $\gamma=0.5$, green line - $\gamma_1=2.5$. Other  parameters are: $\psi_1(0)=\{\frac{1}{\sqrt{2}},0\},\psi_2(0)=\{0,\frac{1}{\sqrt{2}}\},\gamma_0=1, \gamma_2=0.3, \omega_1=\omega_2=\sqrt{37}/2, \varkappa=1, \Delta t=0.001$. The error bars have the same size as the dots on the figure.}\label{Gamma1}
 \end{center}
 \end{figure}

\begin{figure}
   \begin{center}
  \includegraphics[scale=0.5]{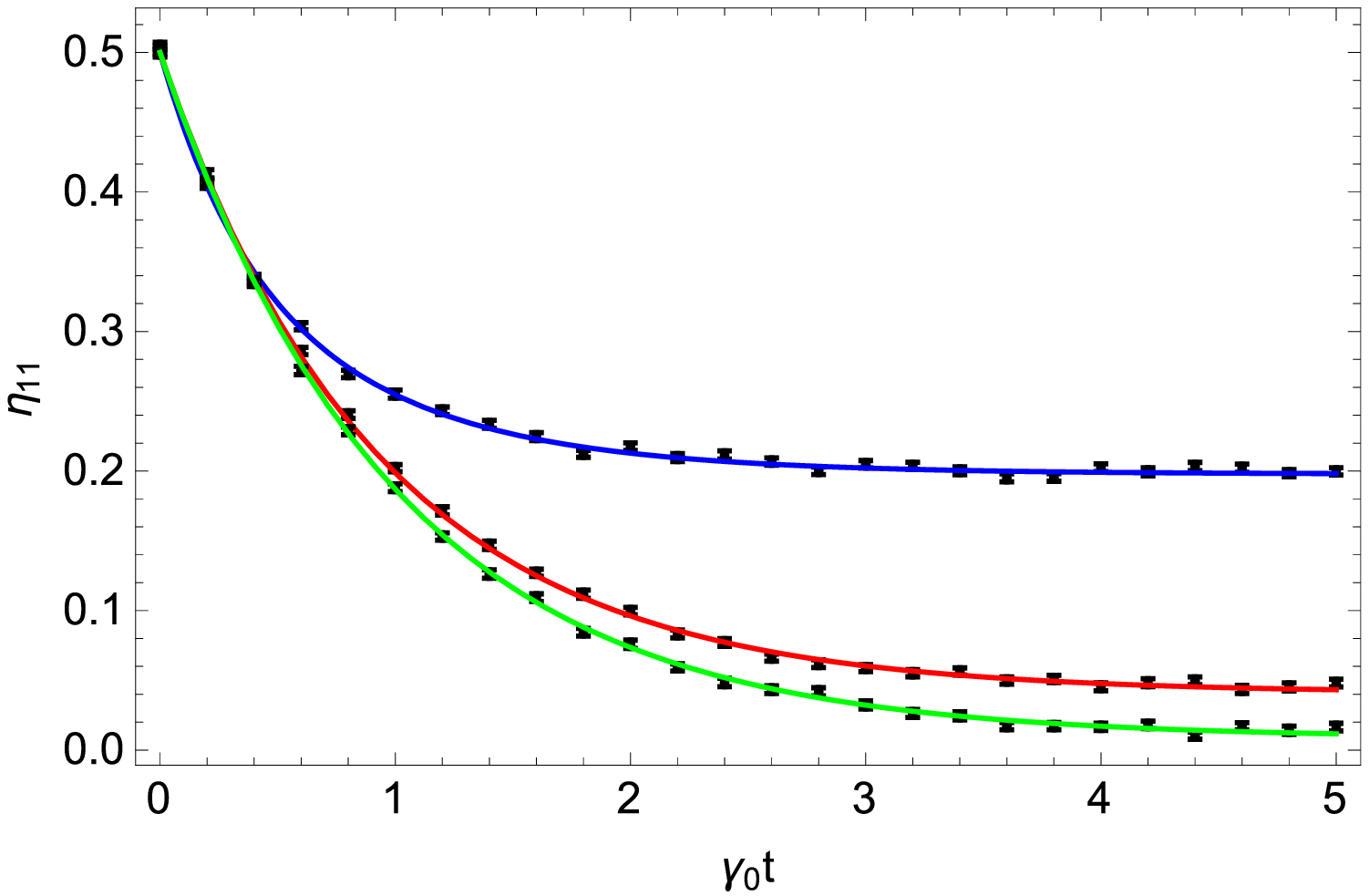}
  \caption{Plot of the mean occupation number of the excited state for different values of $\gamma_2$: red line - $\gamma_2=0.05$, blue line - $\gamma_2=0.3$, green line - $\gamma_2=0.01$. Other parameters are: $\psi_1(0)=\{\frac{1}{\sqrt{2}},0\},\psi_2(0)=\{0,\frac{1}{\sqrt{2}}\},\gamma_0=1, \gamma_1=0.5, \omega_1=\omega_2=\sqrt{37}/2, \varkappa=2, \Delta t=0.001$. The error bars have the same size as the dots on the figure. }\label{Gamma2}
  \end{center}
  \end{figure}

\begin{figure}
  \begin{center}
  \includegraphics[scale=0.5]{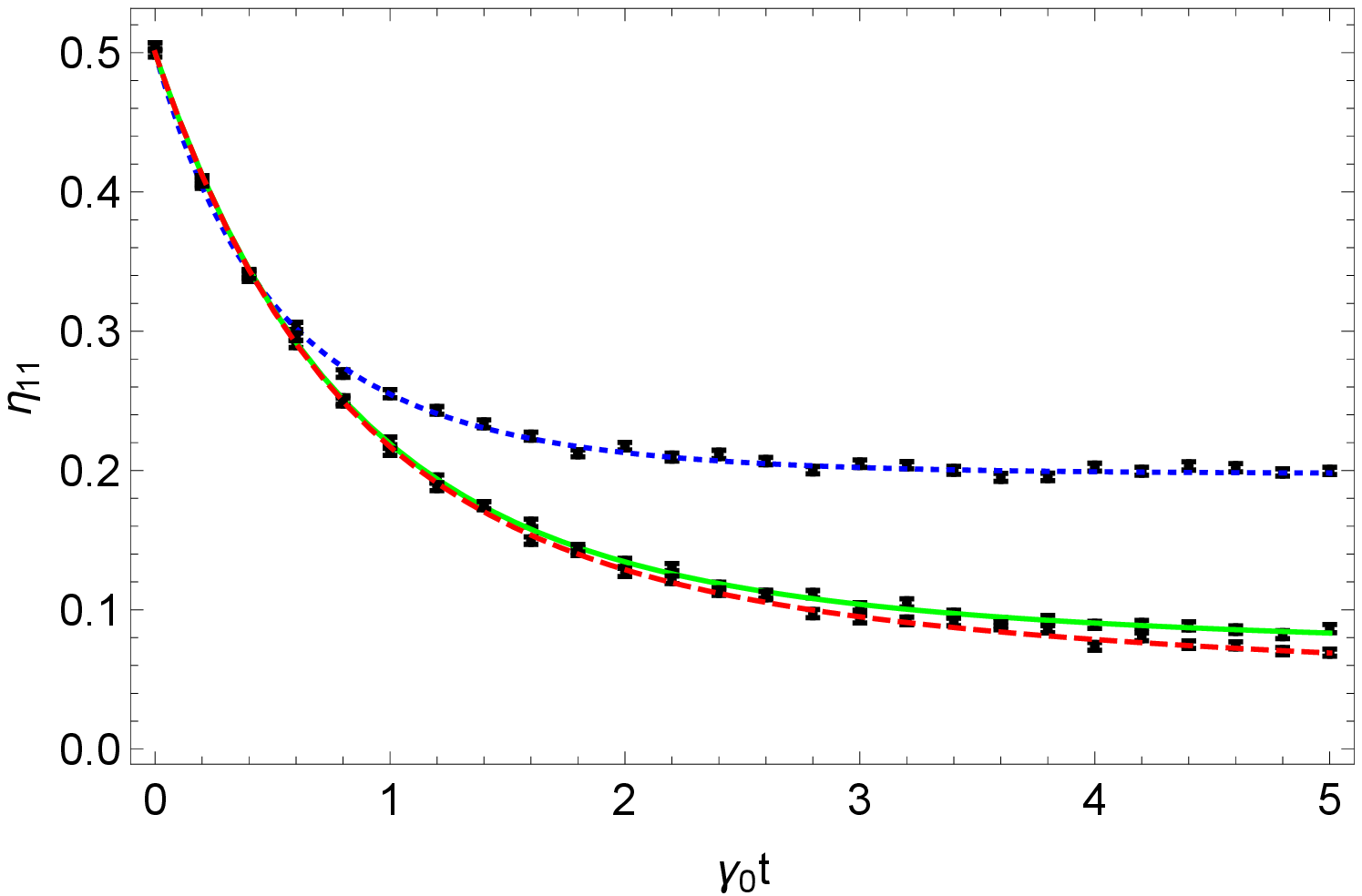}
  \caption{Plot of the mean occupation number of the excited state for different values of $\varkappa$: red line - $\varkappa=0.05$, blue line - $\varkappa=2$, green line - $\varkappa=0.1$. Other parameters are: $\psi_1(0)=\{\frac{1}{\sqrt{2}},0\},\psi_2(0)=\{0,\frac{1}{\sqrt{2}}\},\gamma_0=1, \gamma_1=0.5, \gamma_2=0.3, \omega_1=\omega_2=\sqrt{37}/2, \Delta t=0.001$. The error bars have the same as the dots on the figure.}\label{X}
  \end{center}
  \end{figure}

\begin{figure}[!h]
\begin{center}
\includegraphics[scale=0.5]{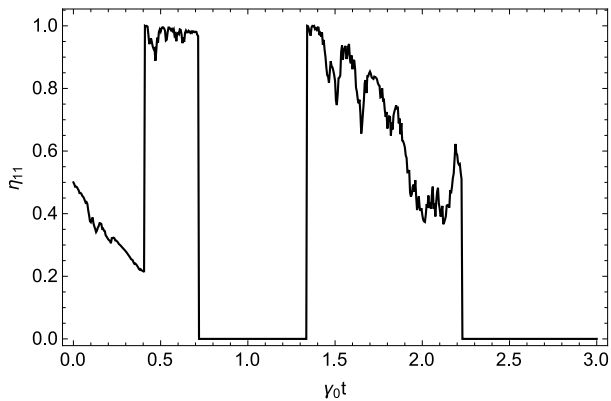}
\caption{A single realization of the process for the parameters: $\psi_1(0)=\{\frac{1}{\sqrt{2}},0\},\psi_2(0)=\{0,\frac{1}{\sqrt{2}}\},\gamma_0=1, \gamma_1=5/2, \gamma_2=3/2, \omega_1=\omega_2=\sqrt{37}/2, \varkappa=1, \Delta t=0.005$.}\label{mod33}
\end{center}
\end{figure}

\begin{figure}[!h]
\begin{center}
\includegraphics[scale=0.5]{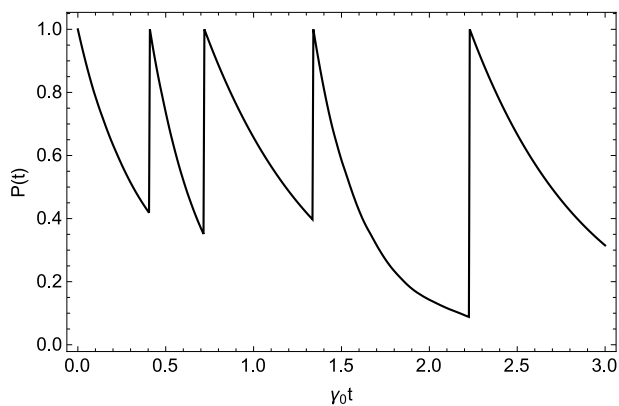}
\caption{The conditional probability of having the next count at time $t$, while the previous count was at time $t_0$. Here $P(t)= $ exp $\{-\int_{0}^t  I(s)ds\}$ after one realization for the parameters: $\psi_1(0)=\{\frac{1}{\sqrt{2}},0\},\psi_2(0)=\{0,\frac{1}{\sqrt{2}}\},\gamma_0=1, \gamma_1=5/2, \gamma_2=3/2, \omega_1=\omega_2=\sqrt{37}/2, \varkappa=1, \Delta t=0.005$.}\label{mod32}
\end{center}
\end{figure}

The results of the simulations using the jump-adapted scheme, discussed above, are shown in Figs. \ref{InStates} - \ref{mod32}. More precisely, all solid lines on the graphs come from the numerical solution of Eq. (\ref{Lindblad}) and the dots demonstrated the Monte Carlo simulations for $10^4$ trajectories. In addition, we show the computational errors overlapping with the dots. 

In Fig. \ref{InStates} the dynamics for the different initial values is shown. The calculations for the mean occupation number of the exited state of the system are implemented by the formula \eqref{M}, that in this case can be rewritten as:
\begin{equation}
\widehat{M}_T=\frac 1 R \sum^{R}_{r=1}\langle\psi^r_T|\sigma_{+}\sigma_{-}|\psi^r_T\rangle,
\end{equation}
 where $R$ is the number of realizations. Starting from different initial conditions, the population tends to the same equilibrium state. This fact is consistent with the intuitive concepts of relaxation. Besides that, the curves corresponding to the exact equation \eqref{Lindblad} are also shown (solid lines on the graphs). It can be seen, that the simulations using the time-dependent rates are in a good agreement with the solution of the corresponding Lindblad rate equation.
 
 Figs. \ref{Gamma1} and \ref{Gamma2} demonstrate the influence of the transition rates $\gamma_1$ and $\gamma_2$. The changing of these parameters affects the dynamics in opposite manner: increasing $\gamma_1$ speeds up the relaxation of the system, while if $\gamma_2$ increases, the relaxation slows down. The different impact of $\gamma_1$ and $\gamma_2$ is primarily connected to its physical value. Besides that, the initial values $\psi_1(0)$ and $\psi_2(0)$ are not equivalent. One also can see an agreement between the numerical values and the exact solutions.
 
As is shown in Fig. \ref{X}, the increasing of $\varkappa$ leads to the acceleration of the relaxation of the system. Again, the numerical results accurately reproduce the exact solution following from Eq. \eqref{Lindblad}. 
 
 A single realization of a trajectory of a jump-diffusive process is shown in Fig. \ref{mod33}. One can see the discrete jumps and the diffusive component between the jumps. The jumps occurs at the moments of time, defined by the conditional probability $P(t)= $ exp $\{-\int_{0}^t  I(s)ds\}$, which is presented in Fig. \ref{mod32}.  The moments of time, when the jumps occurs are clearly seen. For such cases $P(t)=1$. The smoothness of the curve for $P(t)$ can be explained by the averaging of the influence of the diffusion in the integration.
 
\begin{figure}
\begin{center}
\includegraphics[scale=0.5]{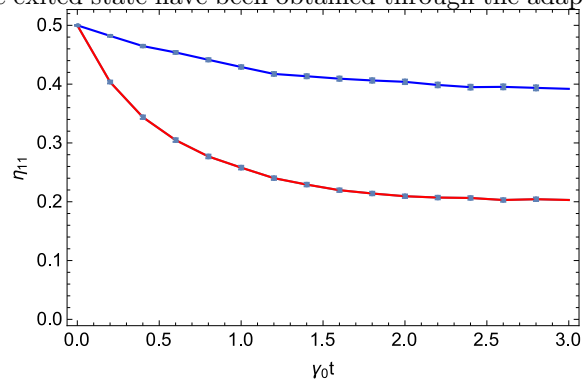}
\caption{Plot of the mean occupation number of the excited state for the parameters $\gamma_1=5/2, \gamma_2=3/2, \omega_1=\omega_2=\sqrt{37}/2, \varkappa=1, \Delta t=0.005$ and $10^5$ realizations. The red line corresponds to the case  with the diffusion in Eq. (\ref{SSE}), while the blue line shows the evolution without diffusion part for the state-dependent intensities. The error bars have the same size as the dots on the figure.}\label{mod34}
\end{center}
\end{figure}

 The last Fig. \ref{mod34} demonstrates the impact of the diffusive component of the process. According to this result, the relaxation slows down without the influence of the diffusion. Physically this corresponds to the absence of an additional relaxation channel associated with the diffusion noise. The simulation results overlap with the results following from Eq. \eqref{Lindblad}.  
 
\section{Discussion}
The generalized Lindblad equation appears in many different situations \cite{1,Bud1, Br1, Bud2, Bud3, Merv,Vacchini, Merv1, Br3, Huang}. All these situations can be unravelled using approach presented in \cite{1} and efficiently simulated by the algorithm presented in this paper. The jump-diffusive unravelling highlights additional relaxation channels that may clarify some details of the quantum evolution. Note that in previous works \cite{Merv1, Huang} only a jump channel of relaxation was considered. 

The generalized Lindblad equation describes non-Markovian relaxation. There are other types of master equations that may be applied to non-Markovian systems, for instance, time-convolutionless master equation (TCL).  It is also possible to construct SSE for the TCL equation \cite{NMTCL}. The last SSE is quite difficult simulate because of all the trajectories has to be evolved simultaneously. In our approach we do not need to evaluate all the trajectories in parallel thanks to the generalized Lindblad form of Eq. \eqref{GME}. 

There are also approaches to non-Markovian quantum systems, based on SSE, which does not require the master equation \cite{3, Sem, 5}. In such approaches the Markovian SSE is generalized by replacing Markovian noise by non-Markovian one. Such SSEs also can be efficiently simulated, but their form is very far from one presented in this paper.

\section{Conclusion}

The jump-diffusion SSEs are a useful tool for the description of non-Markovian evolution. In this paper, we have proposed the numerical approach to simulate these kind of equations. Conceptually, the simulations are performed using the stochastic wave-function method when the quantities of interest are found through the ensemble average of a big number of the realizations of the state vector. This approach is also known as the Monte-Carlo wave-function method.

The model proposed in \cite{1} has been chosen as an object of investigations. It corresponds to the two-level system in a structured bath driven by a multidimensional counting process and multidimensional Brownian motion. The results for the mean occupation number of the exited state have been obtained through the adapted scheme. The curves show a good agreement with the solution of the generalized Lindblad-type master Eq. (\ref{Lindblad}). We have also shown the difference between the curves for the cases, when we have the diffusion and when we turned it off. The presence of the diffusive terms speeds up the dynamics of the system.

\begin{acknowledgments}
This work is based upon research supported by the South African Research Chair Initiative of the Department of Science and Technology and national Research Foundation.
Special thanks to Prof. Alberto Barchielli, whose advices and comments were invaluable.
\end{acknowledgments}


\end{document}